\documentclass[runningheads]{lncse}
\usepackage{multicol} 
\usepackage{makeidx}  
\usepackage{epsfig}

\newcommand{\IGN}[1]{{}}
\makeindex
\begin{document}
\mainmatter              
\addtocmark[1]{Self-Organized Semiconductor Pore Etching} 
\title{Self-Organized Formation of Fractal and Regular Pores in
Semiconductors\protect\footnote{Published in: 
Heike Emmerich, Britta Nestler, and Michael Schreckenberg (Eds.),
Interface and Transport Dynamics - Computational Modelling.
\mbox{Springer (Berlin),}
Lecture Notes in Computer Science and Engineering
{\bf 32}, 82--87 (2003).
}
\vspace*{-3mm}
}
\titlerunning{Self-Organized Pore Formation in Semiconductors}
%
\author{Jens Christian Claussen\inst{1,2}
\and 
J\"urgen Carstensen\inst{1}
\and 
Marc Christophersen\inst{1}
\and 
Sergiu Langa\inst{1}
\and 
 Helmut F\"oll\inst{1}
\vspace*{-3mm}
}
%
\authorrunning{J. C. Claussen, J. Carstensen, M. Christophersen, S. Langa, and H. F\"oll}   
%
\tocauthor{Jens Christian Claussen,J\"urgen Carstensen, Marc Christophersen, 
Sergiu Langa, Helmut F\"oll {\rm(University of Kiel)}}

\institute{
Chair for general materials science, CAU University of Kiel,
Kaiserstr. 2, 24143 Kiel, Germany
\hfill
\texttt{http://www.tf.uni-kiel.de/matwis/amat/}
\and
Theoretische Physik und Astrophysik, Universit\"at Kiel,
Leibnizstr. 15, 24098 Kiel, Germany
\hfill
\texttt{http://www.theo-physik.uni-kiel.de/\~{ }claussen/}
\vspace*{-3mm}
}
\maketitle              
\setcounter{page}{82}
\begin{abstract}
Electrochemical etching of semiconductors, beside
technical applications, provides an interesting experimental setup for
self-or\-ga\-ni\-zed structure formation capable 
of regular, dia\-me\-ter-mo\-du\-la\-ted, and branching pores.
The underlying dynamical processes governing current transfer
and structure formation are described by the Current-Burst-Model: 
all dissolution processes are assumed to occur inhomogeneously in time 
and space as a Current Burst (CB); 
properties and interactions between CB's 
are described by a number of material- and chemistry- dependent ingredients,
like passivation and aging of surfaces in different crystallographic orientations,
giving a qualitative understanding of resulting pore morphologies.
\end{abstract}
\vspace*{-8mm}
\section{Electrochemical Etching: Basic Experimental Setup}
The solid - liquid junction of Silicon and HF - 
containing electrolytes exhibits a number of peculiar features, 
e.g. a very low density of surface states, 
i.e. an extremely well ``passivated'' interface 
\cite{yablonovich86}. 
If the junction is biased, the IV - characteristics 
(Fig.~\ref{fig_kennlinie}) in diluted 
HF is quite complicated and exhibits two current peaks and 
strong current- or voltage oscillations at large current 
densities (for reviews see \cite{foell91appl,smith92}).
These oscillations have been decribed quantitatively by the 
Current-Burst-Model
\cite{carstensen00matsciengb,carstensen98applphys,carstensen98sandiego,carstensen99electrochem}.

The  PC controlled experimental setup is shown in Fig.~\ref{fig_setup}. 
Using a four electrode arrangement, 
a po\-ten\-tio\-stat/gal\-va\-no\-stat 
is contacting the sample and the electrolyte, allowing for a 
well defind potential resp. current for the electrochemical 
dissolution reactions.
Backside contact,
front- and/or backside illumination and electrolyte pumping
can be varied as well as cell size (from under 0.3~cm up to wafers of 6~in),
semiconductor material (Si, InP, GaAs, GaP)
including various doping levels and crystallographic orientations.
In addition, the electrolytes (e.g. HF, HCl, H$_2$SO$_4$) and their 
concentrations and temperature can be varied.
With this setup
a rich variety of porous semiconductors
\cite{propst94,ponomarev98}
can be generated, from nanopores to mesopores and  macropores,
and with various regular and branching structures.

Most of the phenomenonology can be well understood within the 
framework of the Current-Burst-Model which seems to reflect a number of 
quite general properties of semiconductor electrochemistry.

\section{The Current-Burst Model}
\begin{figure}[pthb] \noindent
\hfill
\epsfig{file=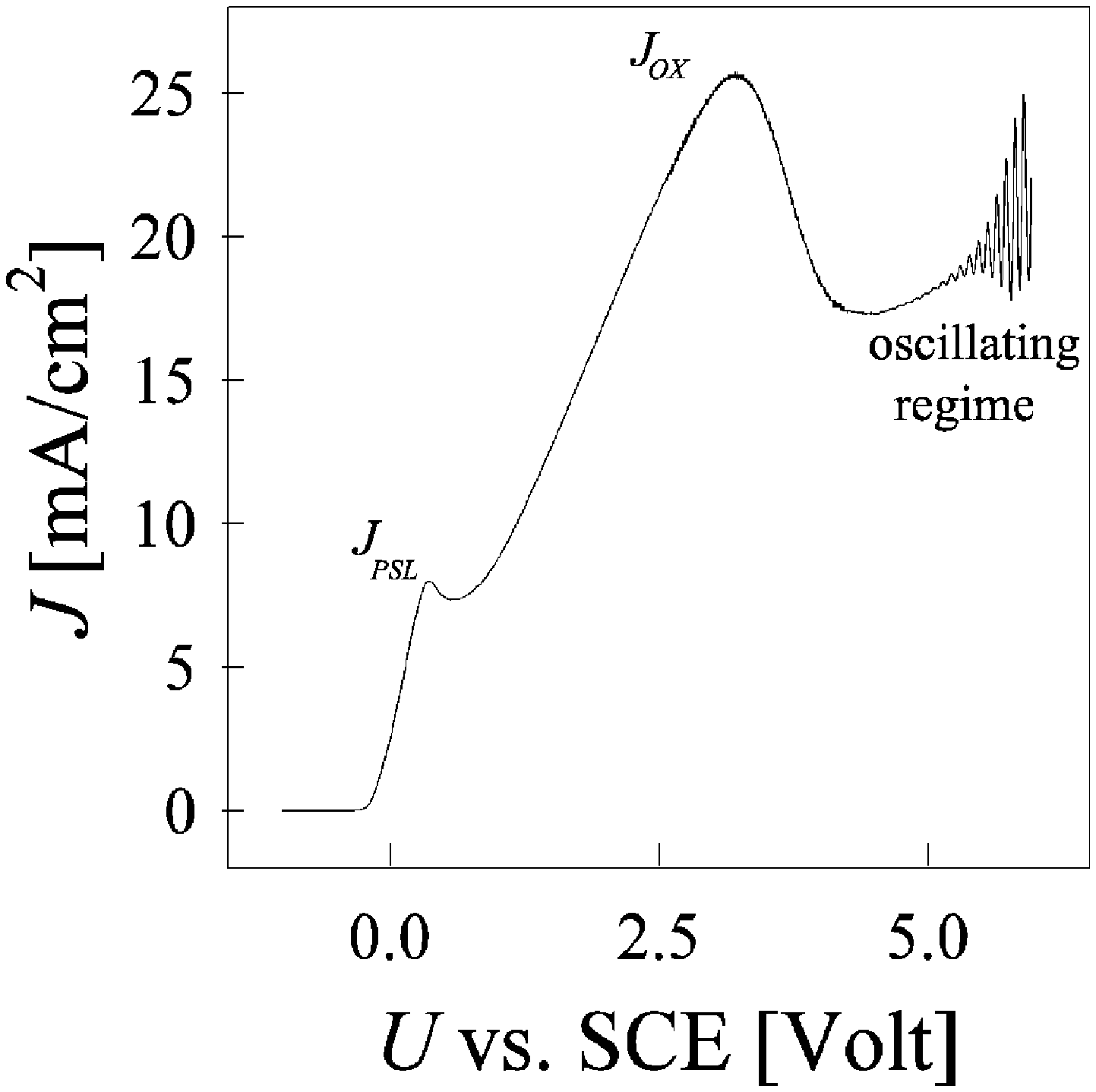,width=0.345\textwidth}
\hfill 
\hfill
\epsfig{file=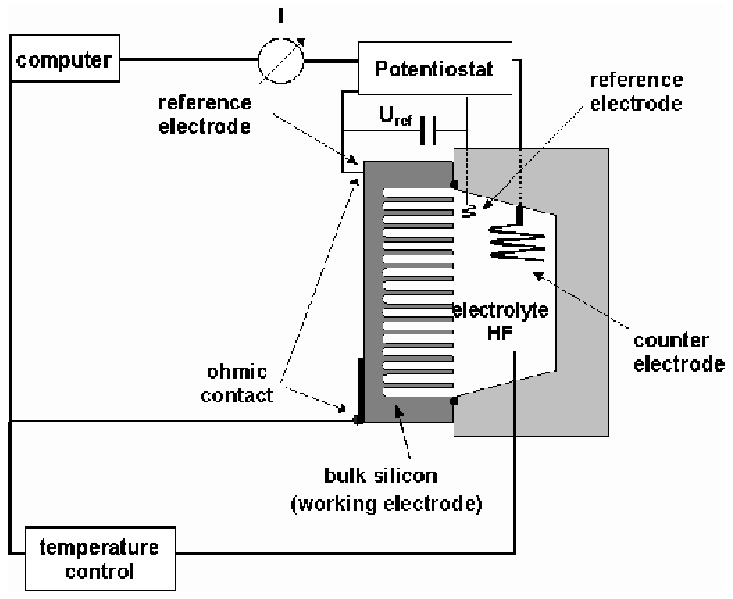,width=0.415\textwidth}
\hfill
\vspace*{-3ex}
\caption{\label{fig_setup}\label{fig_kennlinie}
Left: 
The IV- characteristics of the silicon-hydrofluoric acid contact shows 
different phenomena from generation of a porous silicon layer (PSL), oxidation and
electropolishing (OX) and electrochemical oscillations at higher anodic bias.
Right:
The experimental set-up used for electrochemical anodization of semiconductors.
}
\vspace*{2ex}
\centerline{
\epsfig{file=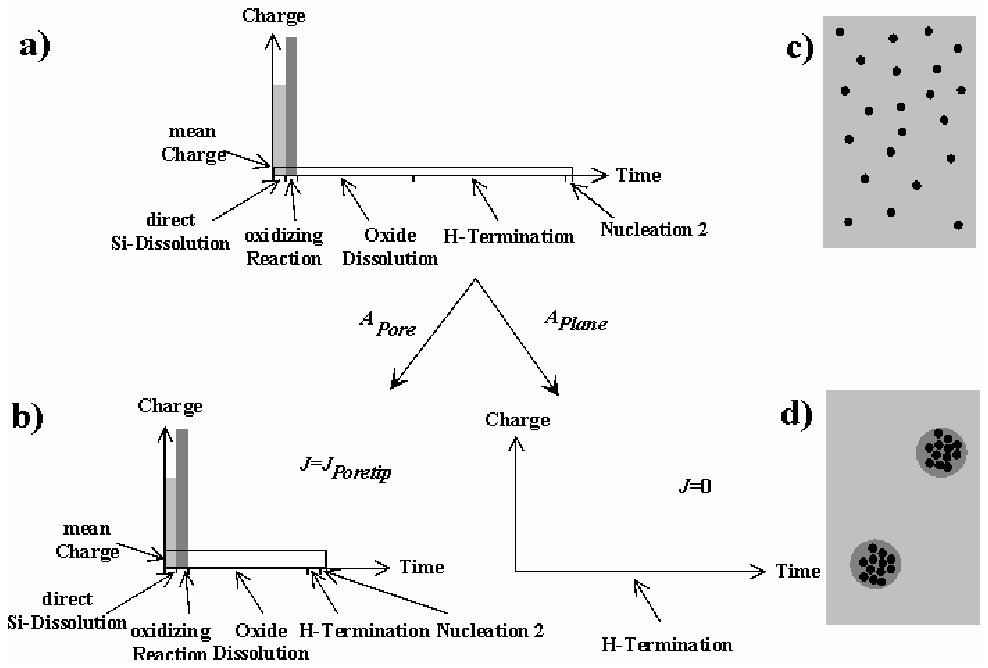,height=0.545\textwidth}
}
\vspace*{-3ex}
\caption{%
\label{fig_schem_charge_time_consuming}%
Applying a global current density smaller than the average current 
density in a current burst will either require very long time constants to 
keep the mean charge density passed in a burst small and to cover the 
surface completely with current lines (a), a statistical arrangement 
of current lines with the optimized (smaller) time constant of the system, 
leaving parts of the surface without current and creating nanopores (b, c), 
or induces a phase separation by rearranging the current lines in areas 
corresponding to macropores (d).
}
\vspace*{2ex}
\centerline{
\epsfig{file=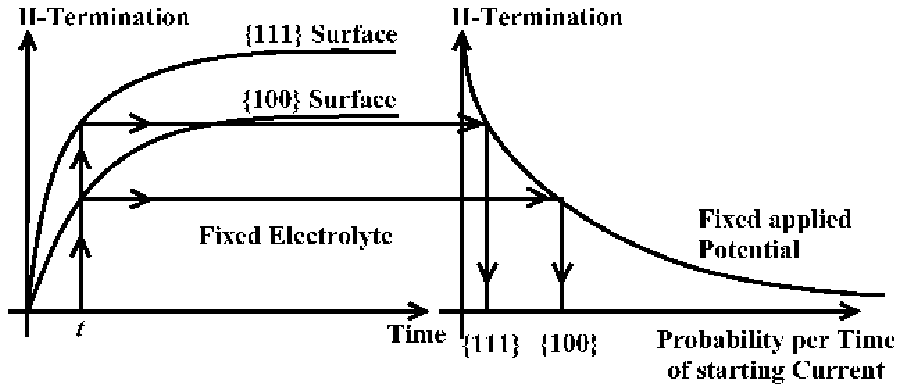,height=0.245\textwidth}
}
\vspace*{-3ex}
\caption{ \label{fig_h_termination}
Time constant and perfection of H-termination differ 
strongly for 
(100) and
(111) Si surface orientation
with  higher nucleation probability on (100) surface.
}
\end{figure}

The Current Burst Model \cite{carstensen00matsciengb,carstensen98applphys,carstensen98sandiego,foellarizona}
states that the dissolution 
mainly takes place on small spots in short events, 
starting with a direct Si-dissolution,
and possibly followed by an oxidizing reaction
(see Fig.~\ref{fig_schem_charge_time_consuming}).

After these two short processes, 
the oxide hump undergoes dissolution,
a time-consuming process which
ensures at the location of the current burst 
a dead-time of fixed length during which no new burst can start.
However, immediately after dissolution the 
Si surface has the highest reactivity,
resulting in a maximal probability of annother
current burst.
Due to H-termination the surface becomes passivated,
and the probability for bursts decays 
until it reaches the properties of a completely
passivated Si surface, comparable to the
situation before the nucleation of the first Current Burst.
\vspace*{-3ex}

\subsubsection{Smoothing effect of Oxide Layers and global current oscillations}
Depending on the regime in the IV-curve 
current bursts and their interaction play
different roles: 
For extremely high currents, in the oscillation
regime, one has a permanent oxide coverage, and 
due to the high forcing of the system
Current Bursts are started at all 
locations where the oxide layer is 
sufficiently thin for a breakthrough.
In this regime, a detailed Monte Carlo study
\cite{carstensen98applphys,carstensen98sandiego,carstensen99electrochem}
including the lateral interaction between
CBs (overlap of the oxide humps of neighboring CBs) 
has shown that the CB Model
quantitatively can explain the experimental 
observations of globally oscillating etching current.
Due to a phase synchronization of neighboring CBs oxide domains are formed. 
The size of this domains increases with increasing oxide generation/reduced oxide dissolution.
At a percolation point only one oxide domain exists on the sample surface with a 
synchronized cycle of oxide growth and dissolution, resulting in a macroscoping 
oscillation of the external current.
The size of the domains as well as the oscillation time can be controlled 
by the chemical parameters. Even without global oscillations for all regions of the IV-curve 
where oxide is formed one finds domains of synchronized oxide growth which define the length 
scale for the roughness of the electrochemically polished sample surface and thus lead 
to a smoothing of the surface. 
\vspace*{-3ex}

\subsubsection{Passivation Effects: The Aging Concept}
While at high current densities the semiconductor surface is completely covered with 
oxide, 
at 
low current 
densities most of the semiconductor surface will be in direct contact to the electrolyte.
It is well known \cite{yablonovich86} that after chemical dissolution the free surface is passivated, 
i.e. the density of surface states reduces as a function of time which increases the stability of the 
surface against further electrochemical attack. Schematically the perfection of the surface passivation 
and the resulting reduction of the probability for a chemical attack as a function of time are plotted 
in Fig. \ref{fig_h_termination}. 
In silicon, the speed and the perfection of 
passivation of the (111) crystallographic surface is larger than for the (100) surface. This selective 
aging of surfaces leads to a self amplifying dissolution of (100) surfaces (which will become pore tips) 
and a preferential passivation of (111) surfaces (which will become pore walls). Under optimized chemical 
conditions with an extremely large passivation difference between (111) and (100) surfaces a self 
organized growth of octahedral cavities occurs.
The octahedra consists of (111) pore walls. As soon as the complete surface of the octahedra reaches a 
critical value, it is easier to start a new cavity at a (100) tip of the old cavity, since the current 
density in the new, small cavity is larger and no surface passivation will occur until the surface again 
becomes to large. This growing mechanism leads to an oscillation of 
both the current through each pore 
and the diameter of each pore as a  function of time.
As in the case of oxide dissolution 
an internal time constant is related to the pore growth.


If a global current density smaller than the average
current density in a current burst is applied, 
the aging concept becomes essential to explain
the generation of macropores: 
At passivated surfaces, the nucleation probability
of new current bursts is lower than at sites 
where a current burst has taken place before.
Therefore the situation shown in
Fig.~\ref{fig_schem_charge_time_consuming}(c)
is much more unlikely than the situation in
Fig.~\ref{fig_schem_charge_time_consuming}(d),
so that the whole surface separates in two
phases with current-carrying pores
and passivated surface without contributions 
to the total current.

%
\section{Consequences on Pore Geometries}
As described above, 
depending on the electrochemical reactions
 the interaction of Current Bursts,
either
surface smoothing due to the  dissolution of an isotropic  oxide layer
or
strongly anisotropic dissolution of the semiconductor 
(due to the crystallography dependent aging of surfaces which is
one of the most important reasons for pore formation) 
will dominate.
Thus one obtains completely 
different surface morphologies when changing the electrochemical 
etching conditions; e.g. in a Si-HF-organic electrolyte system only by 
increasing the HF concentration (i.e. faster oxide dissolution and 
thus reduced influence of oxide) the electrochemical etching 
changes from electropolishing (strong oxide smoothing) 
over macropore formation (pores with diameters of several $\mu$m and 
smooth pore walls) to mesopore formation (strongly anisotropic, narrow 
pores with diameters of less than 400nm) 
\cite{christophersen00physstatsol}.

\begin{figure}[phtb]
\noindent
\epsfig{file=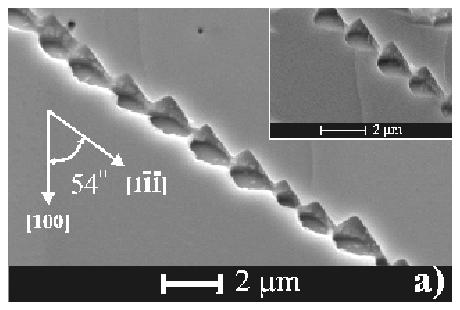,width=0.495\textwidth}
\epsfig{file=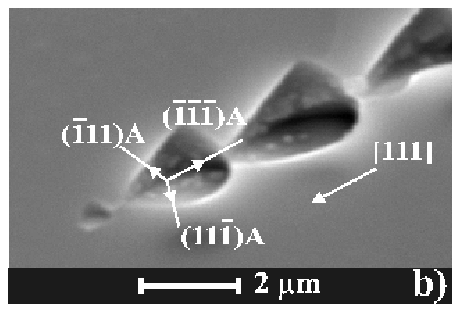,width=0.495\textwidth}
\vspace*{-5ex}
\caption{ \label{fig_tetrahedron}
Tetrahedron-like pores oriented along $<111>$ directions obtained in 
(100)-oriented GaAs at high current densities in HCl electrolytes 
($n$ = 10$^{17}$ cm$^{-3}$).
}
\vspace*{2ex}
\epsfig{file=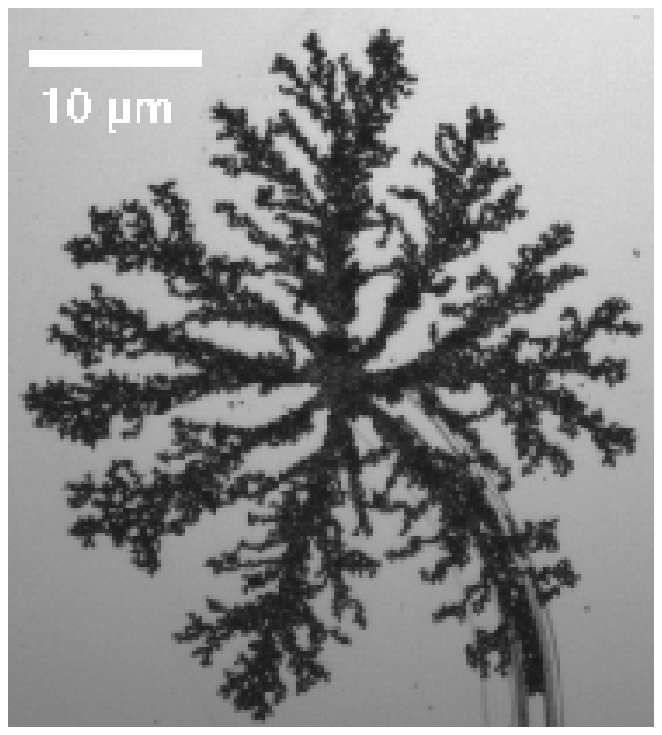, height=0.445\textwidth} \hfill
\epsfig{file=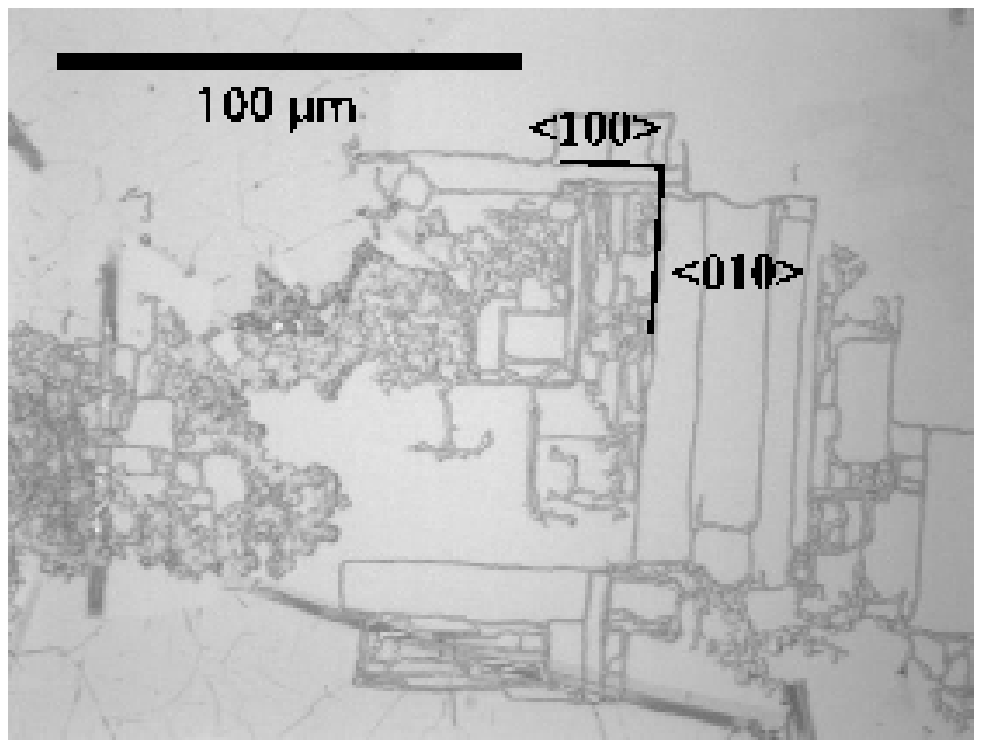,height=0.445\textwidth}
\vspace*{-5ex}
\caption{ \label{fig_fandcrystfract}
Left: Without crystallographic preferentiation,
surface fractals similar to dendritic structures occur.
Right: Under certain etching conditions 
lateral growth along lines in crystallographic
direction occurs which also seems to generate some
self-similar structure.
}
\vspace*{2ex}
\epsfig{file=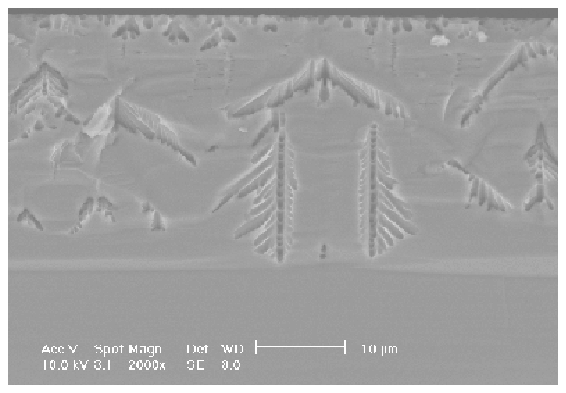,width=0.495\textwidth,angle=0}
\epsfig{file=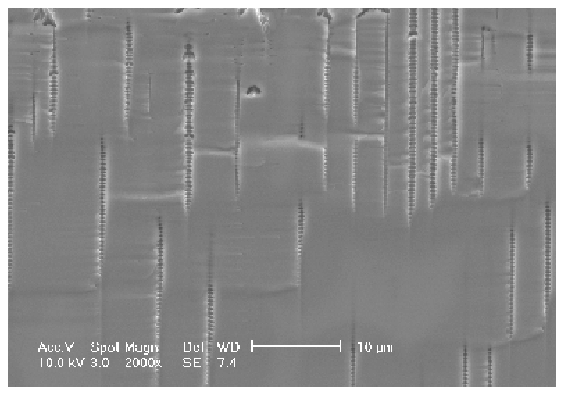,width=0.495\textwidth,angle=0}
\vspace*{-5ex}
\caption{ \label{fig_openloop}
Branching of pores 
(left, for DMF, 4wt-\% HF, constant current)
can be suppressed by 
an open-loop control method triggering to the
system-inherent time scale (right, period 0.5 min).
}
\end{figure}

Silicon has one of the most stable oxides of all semiconductors. So 
generally electrochemical etching of III/V compounds leads
to rougher surfaces and smaller pore diameters. In addition the 
surface aging of III-V compounds is more complicated, since there exist two 
different (111) surfaces; e.g. in GaAs only the $\{111\}$A planes (Ga-rich 
planes) appear as stopping planes. So in most III-V compounds not 
octahedra (eight (111) surfaces as stopping planes) as in silicon are etched 
but tetrahedra with only four (111)A surfaces as stopping planes, 
and the $\{111\}$B planes (As-rich) serve as preferential growing directions 
like the (100) directions in Si (Fig.~\ref{fig_tetrahedron}).
\vspace*{-3ex}

\clearpage

\clearpage
\subsubsection{Lateral Growth of Current Bursts: Generation of
Surface Fractals}
For extremely low current densities and low HF concentration,
the total number of active current bursts is so low that
most of the surface is passivated or covered with oxide. 
CB nucleation becomes a severe problem. 
The neighborhood of a just terminated CB is therefore a preferential site
for new CB nucleation, leading to the formation of surface fractals 
as shown in Fig.~\ref{fig_fandcrystfract} (left). 
A further decrease of the current density
allows for a partial passivation of the surface area of a terminated CB.
According to Fig.~\ref{fig_h_termination} nucleation in the
100 direction of the surface near CBs will be preferential
(Fig.~\ref{fig_fandcrystfract}). 
\vspace*{-3ex}
\subsubsection{Sidebranching Pores and Suppression by Open-Loop-Control}
For technical applications, it would be desirable to
generate pores with a diameter of less than 500nm.
However, at all etching conditions investigated up to 
now, in this diameter range always pores always show
intensive side-branching.
By modulation of the intensity of 
illumination or by modulation of the etching current
with a modulation frequency meet the intrinsic time scale
of the pore formation, 
the side pore formation is hindered 
by the reduction of the etching current at the right phase.
(Fig.~\ref{fig_openloop}). 
\vspace*{-3ex}

\section{Conclusions}
The Current Burst model, a local stochastic
and highly nonlinear model, and the Aging concept 
describing the time- and orientation-dependence
of the passivation behavior,
provide on an intermediate level of abstraction 
a general approach to explain pore geometries,
oscillations and synchronization
with a minimum of material-dependent,
but experimentally accessible, parameters.
The lateral interaction of Current Bursts gives rise to 
synchronization phenomena,
and a percolation transition to global ordering.
Under special conditions lateral growth of 
fractal structures is obtained.
An open-loop control can be successfully applied to suppress
sidebranching of pores. 
To clarify quantitatively the dynamical processes 
in this system, 
further experimental and theoretical
efforts have to be made.
\vspace*{-3ex}

%
%


\begin{thebibliography}{10}
%
\bibitem {yablonovich86} E. Yablonovich, D.L Allara, C.C. Chang, T. Gmitter, T.B. Bright,  Phys. Rev. Lett., 57, 249 (1986)
\bibitem {foell91appl} H. F\"oll, Appl. Phys. A {\bf 53}, 8 (1991) 
\bibitem {smith92} R.L. Smith, S. D. Collins, J. Appl. Phys., {\bf 71}, R1 (1992) 

\bibitem {carstensen00matsciengb}
J. Carstensen,  M. Christophersen, H. F\"oll, 
Mat. Sci. Eng. B, {\bf 69}, 23 (2000)  

\bibitem {carstensen98applphys}
J. Carstensen, R.  Prange, G. Popkirov, H.  F\"oll, 
Appl. Phys. A {\bf 67}, 459 (1998)

\bibitem {carstensen98sandiego}
J. Carstensen, R.  Prange, H. F\"oll,  
in: Proc ECS
San Diego 1998, {\bf 98}  148

\bibitem {carstensen99electrochem} 
J. Carstensen, R. Prange, H.  F\"oll, 
J. Electrochem. Soc. 146(3), (1999) 1134

\bibitem {propst94} E. K. Propst, P.A. Kohl, J. Electrochem. Soc., 141, (1994) 1006
\bibitem {ponomarev98} E. A. Ponomarev, C. Levy-Clement, J. Electrochem. Soc. Lett., 1, (1998) 1002

\bibitem{foellarizona}
H. F\"oll, J. Carstensen, M. Christophersen, G. Hasse, 
in: Proc.
ECS 
2001 

\bibitem{christophersen00physstatsol} 
M. Christophersen, J. Carstensen, H. F\"oll, 
Phys. Stat. Sol. (a) 182, 45  (2000)


\end{thebibliography}
\end{document}